\documentstyle[12pt,epsf]{article}
\begin{document}
\newcommand{\be}{\begin{eqnarray}}
\newcommand{\ee}{\end{eqnarray}}
\newcommand{\jp}{$ \psi $}
\newcommand{\dd}[2]{$ #1 \overline #2 $}
\newcommand\ie {{\it i.e.}}
\newcommand\eg {{\it e.g.}}
\begin{flushright}
LBNL-40680
\end{flushright}
\begin{center}
\vskip 0.2cm
{\Large \bf
$\psi$ Suppression in Pb+Pb Collisions: \\  A New Look at Hadrons vs. Plasma
}
\vskip 0.2cm
R. Vogt$^{\star}$\footnotetext{$^\star$This work was supported
in part by the Director, Office of Energy Research, Division of Nuclear Physics
of the Office of High Energy and Nuclear Physics of the U. S.
Department of Energy under Contract No. DE-AC03-76SF0098.} \\
{\em Nuclear Science Division, Lawrence Berkeley National Laboratory \\
Berkeley, CA, 94720 \\
and \\
Physics Department, University of California at Davis \\
Davis, CA, 95616 }
\end{center}
\vskip 1.0cm
\centerline{\bf Abstract}
\vskip 0.5cm
{\small
A reexamination of hadronic comover scattering
indicates that this mechanism cannot explain
the observed $\psi$ suppression in Pb+Pb interactions.
The possibility of quark-gluon plasma formation is therefore considered.
}
\vskip 1,0cm
\section{Introduction}

Ever since it was realized that the $\psi$ suppression by quark-gluon plasma
formation predicted by Matsui and Satz \cite{MS} could be mimicked by hadronic
means, the source of the observed $\psi$ suppression \cite{na50} has
remained controversial 
\cite{GV2,klns,Wong,Armesto,cassing,capella,plasma}.  In this paper we address
the compatibility of the data with hadronic models in 
light of a reanalysis of our previous work \cite{GV2}.
We also discuss a possible quark-gluon plasma origin of the apparent 
suppression thresholds in the latest Pb+Pb data \cite{na50bnl}.



\section{$A$ Dependence of Hadronic Models}

In this section, we show that the most simplistic assumptions of \jp\ 
suppression
in hadronic interactions lead to the same $A$ dependence for absorption and
secondary, or comover, scatterings.
In hadron-nucleus, $hA$, collisions \cite{GV} 
\be \sigma_{hA \rightarrow \psi} =
\sigma_{hN \rightarrow \psi} \int d^2b \, T_A^{\rm eff}(b) S_A(b) \, \, , 
\label{sigfull} \ee where $b$ is the impact parameter,
$T_A^{\rm eff}(b)$ is the effective nuclear profile, \be T_A^{{\rm
eff}}(b) = \int_{-\infty}^{\infty}\, dz \, \rho_A (b,z) \exp \left\{
-\int_z^{\infty} dz^{\prime} \rho_A (b,z^{\prime}) \sigma_{\psi N}(z^\prime
-z)\right\} \, \, \, , \label{taeff} \ee 
and $\sigma_{\psi N}$ is the cross section
for $\psi$ (or $|c \overline c g \rangle$) absorption by nucleons.   
The nuclear density distributions, $\rho_A(b,z)$, are taken from nuclear charge
density measurements \cite{JVV}.
The comover survival probability is
\cite{GV} \be S_A  \approx \exp \left\{ -
\int_{\tau_0}^{\tau_I}\!d\tau\,{\langle\sigma_{\psi {\rm co}} 
v\rangle}n(\tau, b)\right\} = \exp \left\{ - {\langle\sigma_{\psi {\rm co}} 
v\rangle} n_0 \tau_0 \ln\left( \frac{\tau_I}{\tau_0} \right) \right\}
\, \, , \label{comosurv} \ee
where $\sigma_{\psi {\rm co}}$ is the $\psi$--comover 
absorption cross section, 
$v \approx 0.6$ is the relative $\psi$-comover velocity, and
$n(\tau,b)$ is the density of comovers at time $\tau$ and impact parameter $b$.
Integrating over $\tau$ and relating the initial density to
the final hadron rapidity density \cite{Bjorken},
$n_0 \tau_0 = (\pi R_h^2)^{-1} (dN/dy)$,
we have \cite{GV} \be \langle\sigma_{\psi {\rm co}} v\rangle \int d\tau \, 
n(\tau, b) \approx \frac{\langle\sigma_{\psi {\rm co}} v\rangle}{\pi R_h^2}
\sigma_{hN} T_A(b)\frac{dN}{dy}|_{y_{\rm cm}} 
\ln\left( \frac{\tau_I}{\tau_0} \right) \, \, , \label{comofull}
\ee where $dN/dy|_{y_{\rm cm}}$, the central rapidity 
density in an $hp$ collision, is scaled up to $hA$ interactions by
$\sigma_{hN}T_A(b)$, the number of participants.
Since $\pi R_h^2 \approx \sigma_{hN}$, these two factors cancel.
The effective proper lifetime $\tau_I$ over which the comovers, formed at time
$\tau_0$, interact with the $\psi$ is $\tau_I \sim R_h/v$.

Expanding the exponentials in eqs.~(\ref{taeff}) and (\ref{comosurv})
to terms linear in 
cross section, integrating eq.~(\ref{sigfull}) and re-exponentiating, we
obtain \cite{GV} 
\be \frac{\sigma_{hA \rightarrow \psi}}{A \sigma_{hN
\rightarrow \psi}} & \approx & \exp \left\{ - \frac{9 A^{1/3}}{16 \pi
r_0^2}\left[ \sigma_{\psi N} + 2 \langle \sigma_{\psi {\rm co}} v \rangle
\ln \left( \frac{\tau_I}{\tau_0} \right) \frac{dN}{dy}|_{y_{\rm cm}} \right] 
\right\} \\ & = & \exp \left\{ -A^{1/3} \left( \eta + \beta \right) \right\} 
\, \, . \label{psiha} \ee For large targets with $A>50$, $A^{1/3} \approx \ln 
A$ so that \be \frac{\sigma_{hA \rightarrow \psi}}{\sigma_{hN
\rightarrow \psi}} = A^{1-\eta - \beta} = A^\alpha \, \, . \label{psinalf} \ee
The comover contribution to $\alpha$ 
could be significant \cite{GV} unless $\sigma_{\psi {\rm co}}$ is 
strongly reduced at low 
energies \cite{kssig} (although see also \cite{KSSZ}),
$\tau_I$ is very short or $dN/dy$ is small. In most recent analyses 
\cite{GV2,klns,Armesto,cassing,GSTV}, an effective value of $\sigma_{\psi N}$
including both effects is determined from $\alpha$.

Unfortunately the identical nature of the two contributions to 
eq.~(\ref{psinalf}) suggests that they
may be inextricably intertwined. Experimental constraints on $\sigma_{\psi N}$
and $\sigma_{\psi \rm co}$ would clearly be valuable.
Exclusive \jp\ production in near threshold $\overline p A$ interactions
\cite{kseth} and an inverse kinematics experiment \cite{KSinv}
could clarify $\sigma_{\psi N}$. 
A measurement of $dN/dy$ associated with \jp\ production could place limits 
on the comover contribution. 

\section{$E_T$ Dependence of Hadronic Models}

\begin{figure}[htb]
\setlength{\epsfxsize=0.95\textwidth}
\setlength{\epsfysize=0.35\textheight}
\centerline{\epsffile{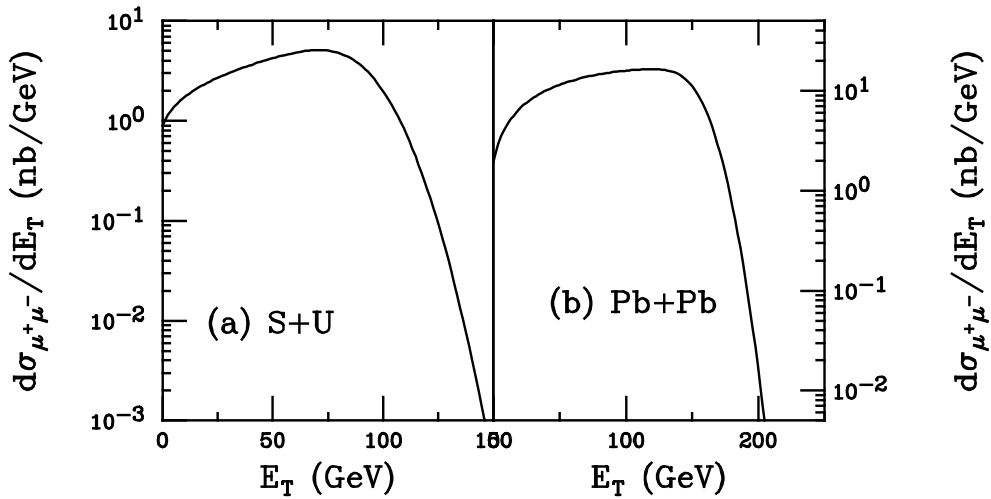}}
\caption[]{The $E_T$ distributions of Drell-Yan pairs with $2.9<M<4.5$ GeV
in (a) S+U and (b) Pb+Pb
interactions.  The $pp$ normalization is calculated 
at 200 GeV with the GRV LO \cite{GRV} distributions and
$|\cos \theta_{\rm CS}|<0.5$.}
\label{etdists}
\end{figure}

The transverse energy, $E_T$, dependence of \jp\ production in
nucleus-nucleus collisions is generally compared to Drell-Yan, DY, production, 
uninfluenced by final-state interactions.
The $\psi$/DY ratio as a function of $E_T$ is
\be \frac{B\sigma_{\psi}}{\sigma_{\mu^+ \mu^-}}(E_T) = 
\frac{B \sigma_{pp \rightarrow
\psi} \int\, d^2b \int\, d^2s \, T_{A}^{\rm eff}(b)\, T_{B}^{\rm eff}(|\vec b
- \vec s|)\, S_{AB}(b,s) \, p(E_T;b)}{\sigma_{pp \rightarrow
\mu^+ \mu^-} \int\, d^2b \int\, d^2s \, T_{A}(b)\, T_{B}(|\vec b
- \vec s|)\, p(E_T;b)} \label{psiet} \, \, . \ee
The probability to produce $E_T$ at impact parameter $b$, $p(E_T;b)$, 
is a Gaussian with mean $\overline E_T(b) = \epsilon_N N_{AB}(b)
$ proportional to the number of participants and standard deviation $\sigma^2
(b) = \omega \epsilon_N \overline E_T(b)$ \cite{GV}.  The parameters 
$\epsilon_N$, the energy per participant, and $\omega$ are chosen
to agree with the NA38/NA50 neutral $E_T$ distributions.  In S+U interactions 
with  $1.7<\eta<4.1$ \cite{na38}, $\epsilon_N
= 0.74$ GeV while in Pb+Pb interactions with $1.1<\eta<2.3$ \cite{na50}, 
$\epsilon_N = 0.4$ GeV.
We use $\omega = 3.2$ but note that the precise value of $\omega$ only weakly
influences the $\psi$/DY ratios.
In Fig.~\ref{etdists} the Drell-Yan $E_T$ distributions are shown 
normalized to the $pp$ cross section at 200 GeV.  (See also \cite{Dias}.)

The $\psi$, $\psi'$ and $\chi_c$ are assumed to
interact with nucleons while in  $|c \overline c g \rangle$
color octet states with a lifetime of $\approx 0.3$ fm \cite{KSoct}.  
Because the final charmonium state has formed by the time it interacts with
comovers, the $\sim 30$\% $\chi_c$  \cite{Teva} and 
$\sim 12$\% $\psi'$ \cite{HPC} decay contributions to \jp\ production
are considered separately. 
To include $\psi'$ and $\chi_c$ feeddown to the $\psi$, 
we scale the $\psi'$ and $\chi_c$ comover cross sections relative to
$\sigma_{\psi {\rm co}}$ by the squares of the
radii \cite{hpov} so that 
$\sigma_{\psi' \rm co} \approx 3.8 \sigma_{\psi \rm co}$,
$\sigma_{\chi_c \rm co} \approx 2.4 \sigma_{\psi \rm co}$ and
\be S_{AB}(b,s) = 0.58 \, S_\psi(b,s) + 0.3 \, S_{\chi_c}(b,s) + 0.12 \,
S_{\psi'}(b,s) \, \, . \label{survfeed} \ee
In Refs.\ \cite{GV2,GSTV} the density of comovers was assumed to
be directly proportional to $E_T$, $n_{\rm co} = {\overline
n} E_{T}/{\overline E}_{T}(0)$ with $\tau_0 \approx 2$ fm and 
$\tau_I \approx R_A/v$.  However, since $E_T$ is proportional to the number of
participants, we replace $n_0 \tau_0$ by $n_{AB}(b,s)$,
the participant density \cite{plasma},
\be S_{AB}(b,s) = \exp \left\{- \langle \sigma_{\rm
co} v \rangle n_{AB}(b,s)
\ln \left(\frac{\tau_I(b)}{\tau_0(b,s)} \right) \right\} \, \, , 
\label{comoabfull} \ee
similar to Ref. \cite{klns}.
The comover formation time, a function of the path length $L$ \cite{gvpt,gh},
$\tau_0(b,s) = 1 + L_A(b,s)/\gamma_A(b,s) + 
L_B(b,s)/\gamma_B(b,s)$, 
is $\sim 2$ fm in central collisions and $\sim 1$ fm in very peripheral
collisions. The comovers interact with the \jp\ only if $\tau_I(b) > 
\tau_0(b,s)$ where
\be \tau_I(b) = \left\{ \begin{array}{ll} R_{A}/v \,\, \, \, \, \,
\, & \mbox{$b<R_{B} -R_{A}$} \\ (R_{A} + R_{B} - b)/(2v) &
\mbox{$R_{B} - R_{A} < b< R_{B} + R_{A}$} \end{array}     \right. \, \,
. \ee

\begin{figure}[htb]
\setlength{\epsfxsize=0.95\textwidth}
\setlength{\epsfysize=0.4\textheight}
\centerline{\epsffile{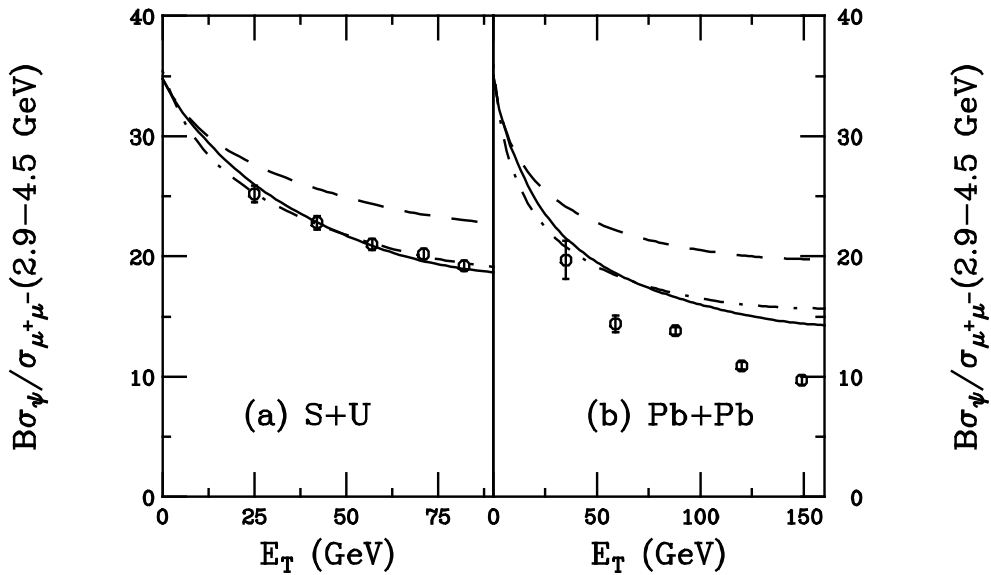}}
\caption[]{The $\psi$/DY calculations are compared with data \cite{na50}
from S+U (a) and Pb+Pb (b)
interactions, both at 200 GeV. The solid curve shows $\sigma_{\psi N} = 4.8$ mb
and $\sigma_{\psi {\rm co}} = 0.67$ mb.  The
dashed and dot-dashed assume $\sigma_{\psi {\rm co}}=0$ for 
$\sigma_{\psi N} = 4.8$ mb and $\sigma_{\psi N} = 7.3$ mb, respectively.}
\label{etrat}
\end{figure}

Figure~\ref{etrat} shows the $\psi/$DY ratios for  $\sigma_{\psi N} =
7.3$ mb and 4.8 mb. 
Comovers are included when $\sigma_{\psi N} =
4.8$ mb.  Previously \cite{GV2,GSTV} we used $\sigma_{\psi {\rm co}} 
\approx 2\sigma_{\psi N}/3 = 3.2$ mb from quark-counting arguments; here we
find $\sigma_{\psi {\rm co}} = 0.67$ mb fits the S+U data, similar to the
results of Ref.~\cite{Armesto}.  
When feeddown is not included,  $\sigma_{\psi {\rm co}} = 1.1$ mb produces 
equivalent agreement. The \jp\ cross 
section in $pp$ interactions is fit to the $A$ dependence given each value of 
$\sigma_{\psi N}$: $\sigma_{pp \rightarrow \psi} =
2.1$ nb for $\sigma_{\psi N}= 4.8$ mb and  $\sigma_{pp \rightarrow \psi} =
2.3$ nb for $\sigma_{\psi N} = 7.3$ mb.  Assuming the maximum
comover $pA$ contribution when $\sigma_{\psi N}= 4.8$ mb is equivalent to an
effective $\sigma_{\psi N}$, eqs.~(\ref{psiha},\ref{psinalf}), 
of $\approx 6$ mb, in
agreement with the $pA$ data \cite{na50}.  The values of $\sigma_{pp 
\rightarrow \psi}$ are in agreement with both the data \cite{na50} and 
low energy parameterizations of the cross section
\cite{cassing,Schuler} as well as with next-to-leading order calculations
\cite{HPC}.  

The results in Fig.~\ref{etrat} differ from those obtained in Ref.~\cite{GV2}
where simultaneous agreement of the model with both the S+U and Pb+Pb data was
found. The agreement 
with the NA38 S+U data \cite{na50}
has improved over that of Refs.~\cite{GV2,GSTV}. 
However, the result now disagrees with the Pb+Pb data, in accord with the
conclusions of Refs. \cite{klns,Wong}.  The major
difference lies in the normalization of the $\psi/$DY ratio,
determined from the $pp$ production cross sections in the
NA50 phase space.  All the data \cite{na50,na38,na51} 
has been isospin adjusted to 
the $pp$ cross section, continuum mass to $2.9<M<4.5$ GeV, 
projectile energy to 200 GeV,  and phase space to
$0<y_{\rm cm}< 1$ and  $|\cos \theta_{\rm CS}|<0.5$.  The leading order
Drell-Yan cross 
sections were adjusted for mass and isospin using the GRV LO \cite{GRV}
distributions with the NA50 $K$ factor \cite{na50}.  The isospin correction 
at $\sqrt{s} = 19.4$ GeV and $2.9<M<4.5$ GeV is $pp$/Pb+Pb $\approx 
1.3$.  Since $\psi$ production is
dominated by gluon fusion, it is insensitive to isospin.  In 
Refs.~\cite{GV2,GV,GSTV} the S+U data was compared to the ``continuum'' mass
range, $1.7 < M < 2.7$ GeV, and extrapolated to $|\cos \theta_{\rm CS}|<1$.  
Contrary to what was stated in Ref.~\cite{GV2}, the angular adjustment from 
$|\cos \theta_{\rm CS}|<1$ to $|\cos \theta_{\rm 
CS}|<0.5$ was left out of the Pb+Pb calculation.  Thus a 23\% increase of the
normalization in Ref.~\cite{GV2} is needed because while
the \jp\ decays decays isotropically to leptons \cite{angdep}, requiring a 
factor of two adjustment,
the Drell-Yan cross section, $\propto 1 + \cos^2 
\theta_{\rm CS}$ \cite{hpdy}, is adjusted by 2.46.
Therefore, although assuming $\sigma_{\psi N}
= 4.8$ mb still leads to the earlier conclusion 
that comovers are necessary to explain the S+U data \cite{GV,GSTV}, the Pb+Pb
results now suggest that the more absorption is needed.

Other recent comover calculations \cite{Armesto,cassing}
find simultaneous agreement between the two systems within 
dynamical models of secondary production, consistent with scaling $n_{\rm
Pb+Pb}$ by a factor of two.  If
secondary production in the central rapidity region increased $\propto 
n_{AB}^x$ where $x>1$, as suggested by recent
results from NA49 \cite{na49}, the discrepancy between our results and the
Pb+Pb data could be reduced at high $E_T$, {\it e.g.} if $x=1.3$, $\psi$/DY
decreases 13\% at $E_T=150$ GeV.
Agreement of the comover interpretation with the data
was also found in Ref.~\cite{capella}.  While some of the differences with our
conclusions lie in the choice of model parameters and
the specific comover $E_T$ dependence, the results are
also rather sensitive to the nuclear density profile.  The S+U $p_T$-dependent
data favor a Woods-Saxon density profile over a sharp surface nucleus 
\cite{gvpt}.  However, the precise Woods-Saxon may also affect the
agreement of the model with the data. For example,
Ref.~\cite{Armesto} uses a profile with lower central nuclear densities
\cite{nestor} than those determined from data \cite{JVV}.  
In particular, $\rho_{\rm S}(0,0)$
is $\approx 35$\% lower while $\rho_{\rm U}(0,0)$ and $\rho_{\rm Pb}(0,0)$ are
$\approx 10$\% lower.

\begin{figure}[htb]
\setlength{\epsfxsize=0.95\textwidth}
\setlength{\epsfysize=0.4\textheight}
\centerline{\epsffile{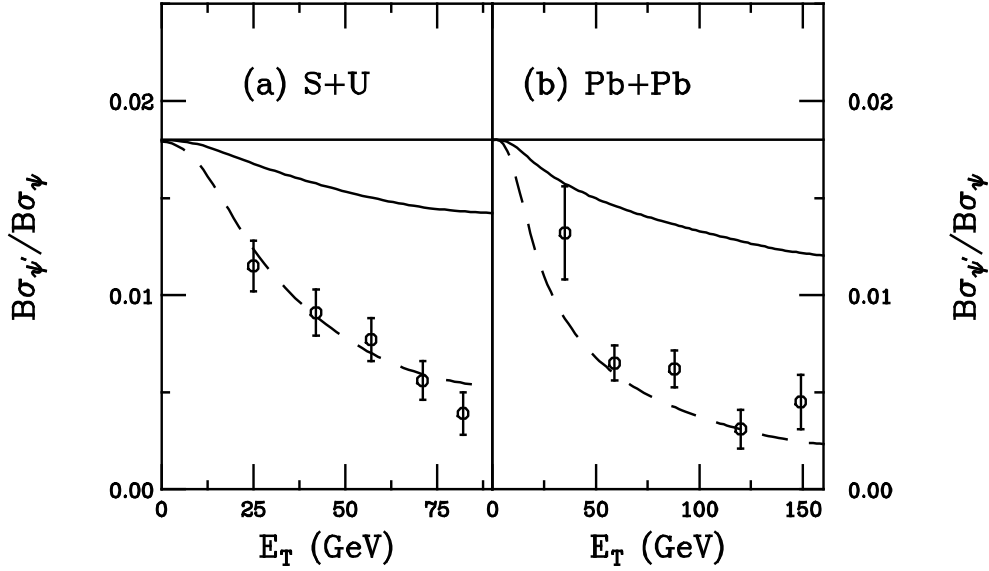}}
\caption[]{The calculated $\psi'/\psi$ ratio is compared with data from
S+U (a) and Pb+Pb (b) collisions \cite{na50ind}. 
The solid curves show $\sigma_{\psi N} = \sigma_{\psi' N} =
4.8$ mb, $\sigma_{\psi {\rm co}} = 0.67$ mb and $\sigma_{\psi' \rm co} = 3.8
\sigma_{\psi \rm co}$. The dashed curves, neglecting feeddown,
take $\sigma_{\psi' {\rm co}} = 8.5$ mb.}
\label{etratpp}
\end{figure}

The $\psi'/\psi$ ratios \cite{na50ind} are shown in Fig.~\ref{etratpp}.  
If $\sigma_{\psi' \rm
co}$ is a free parameter, neglecting feeddown, then  
$\sigma_{\psi' {\rm co}} \approx 7$-8.5 mb
reproduces the observed suppression for $\sigma_{\psi N} =
7.3$ mb with $\sigma_{\psi \rm co} \approx 0$ and $\sigma_{\psi N} = 4.8$ mb,
$\sigma_{\psi \rm co} = 0.67$ mb respectively.
Including feeddown, eq.~(\ref{survfeed}),
$\sigma_{\psi' \rm co} \propto \sigma_{\psi {\rm co}}$
and the calculated ratios underpredict the data.
Other combinations of $\sigma_{\psi \rm
co}$, $\sigma_{\chi_c \rm co}$ and $\sigma_{\psi' \rm co}$ produce
similar $\psi/$DY results but do not significantly improve agreement with the
$\psi'/\psi$ ratios.

\section{Suppression by Plasma Screening?}

In light of these hadronic results,
we speculate as to the nature of $\psi$ suppression by
plasma screening.  The quarkonium potential is expected to be
modified at finite temperatures by the screening mass
$\mu(T)$. When $\mu(T) \geq \mu_D$, screening 
prevents resonance formation at $T=T_D$.  The critical
values are $\mu_D^{\chi_c}= 0.342$ GeV, $\mu_D^{\psi'} = 0.357 $ GeV and
$\mu_D^\psi = 0.699 $ GeV \cite{KMS}.  Whether a simple plasma interpretation
is possible depends strongly on $\mu(T)$.
Plasma suppression is possible when the
energy density, $\epsilon$, is greater than $\epsilon_D$.

We assume that $\mu(T) = \sqrt{1 + n_f/6} \, g(T) T$
where $g^2 (T) = 48\pi^2/[(33-2n_f) \ln F^2]$ with
$F = K (T/T_c)
(T_c/\Lambda_{\overline{MS}})$ \cite{lh}.
In SU(3) gauge theory, $T_c = 260$ MeV \cite{boyd} and $T_c/\Lambda_{\rm 
\overline{MS}} = 1.03 \pm 0.19$ \cite{fingberg}. A fit to the heavy quark
potential at high temperatures yields $K \approx 33.8$ \cite{lh}.
Lattice results with $n_f =$2 and 4 suggest $T_c = 170$ MeV and 
$T_c/\Lambda_{\rm \overline{MS}} = 1.05$ \cite{fingberg}.  
If the SU(3) value of $K$ is applicable when $n_f>0$, then for 
$n_f = 2$ or 3, the
$\chi_c$ breaks up at $T_D^{\chi_c}\approx 180$-200 MeV and
$\epsilon_D^{\chi_c} = 2.1$-2.6 GeV/fm$^3$ while the
$\psi'$ subsequently breaks up at $T_D^{\psi'}\approx 190$-210 MeV
and $\epsilon_D^{\psi'} = 2.6$-3.1 GeV/fm$^3$.  
The $\psi$ itself would not break up until $T>2T_c$.
If $n_f = 4$, $T_D = T_c$ for both the $\chi_c$ and $\psi'$ with
$\epsilon_D = \epsilon_c \approx 2.1$ GeV/fm$^3$ while
$T_D^\psi \approx 370$ MeV.  The $n_f = 0$ results are similar to those with
$n_f = 4$ albeit with correspondingly larger $\epsilon$ due to the higher
$T_c$.  However, the high temperature limit
is probably invalid for $T/T_c \leq 3.5$ \cite{heller}.  A fit of $K$ to
lattice data for $T \geq T_c$ \cite{lh} results in larger values of $\mu(T)$, 
suggesting $T_D =
T_c$ for the $\chi_c$, $\psi'$ and $\psi$ with $\epsilon_D = \epsilon_c 
\approx 1.3$-2.1 GeV/fm$^3$ for $n_f = 2$-4 and $\epsilon_D = \epsilon_c
\approx 3.1$ GeV/fm$^3$ for $n_f = 0$.

Thus the $\psi/$DY ratio
could contain one or two thresholds, as suggested by the new Pb+Pb data
\cite{na50bnl}.  We choose three illustrative cases: I) $n_f =3$,
sequential $\chi_c$ and $\psi'$ break up; II) $n_f = 4$, $T_D^{\chi_c} =
T_D^{\psi'} = T_c$; and III) $T_D = T_c$ for all charmonium states.  Cases I
and II assume $K \approx 33.8$ while case III takes $K$ from
the fit for $T \geq T_c$ \cite{lh}. 

\begin{figure}[htb]
\setlength{\epsfxsize=0.95\textwidth}
\setlength{\epsfysize=0.4\textheight}
\centerline{\epsffile{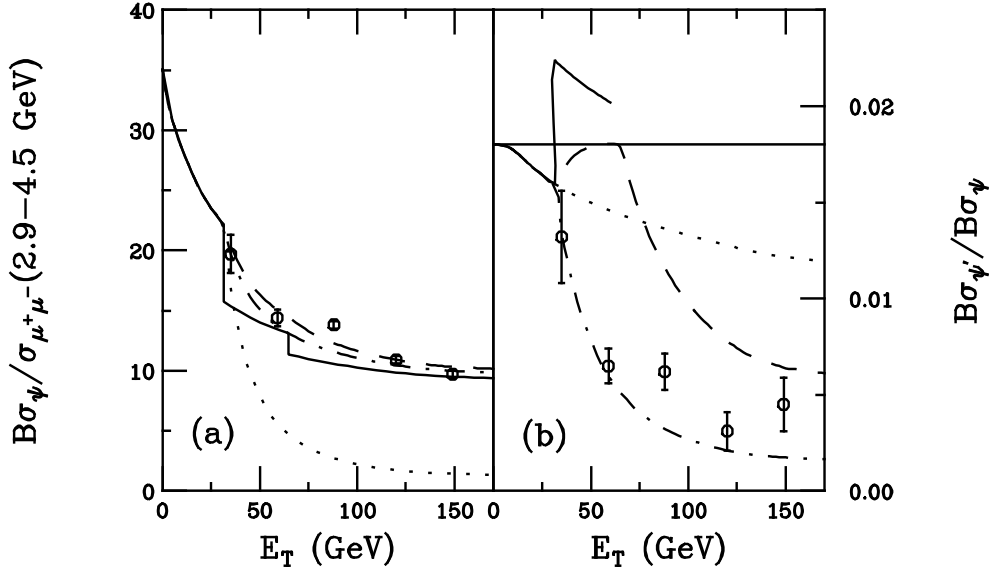}}
\caption[]{ The Pb+Pb $\psi/$DY \cite{na50} (a) and 
$\psi'/\psi$ \cite{na50ind} (b) ratios at 200 GeV compared to 
plasma predictions assuming $\sigma_{\psi N} = \sigma_{\psi' 
N} = 4.8$ mb, $\sigma_{\psi {\rm co}} = 0.67$ mb and $\sigma_{\psi' {\rm co}} 
= 3.8 \sigma_{\psi \rm co}$.  The solid and
dashed curves illustrate case I with total suppression above $\epsilon_D$ and
with $R=R_{\rm Pb}$ respectively.  The dot-dashed and dotted curves correspond
to cases II and III with $R=R_{\rm Pb}$. 
}
\label{figuresepssun}
\end{figure}

The relationship between $\epsilon$,
not directly measurable, and $E_T$ is determined from $n(E_T)$,
the average number of subcollisions per area \cite{klns,plasma,gvpt}.
Using the NA50 prescription \cite{bernard} to calculate $\epsilon$
in Pb+Pb collisions places the maximum $\epsilon$ in S+U collisions below
the first threshold in the Pb+Pb data \cite{na50bnl}.

If suppression is assumed to be total at $\epsilon =
\epsilon_D$, then sharp thresholds appear, as shown
for case I in  Fig.~\ref{figuresepssun}(a).  Such sharp thresholds seem
physically unlikely due to the fluctuations in $E_T$, the finite size of the
plasma, and the finite momentum of the charmonium state.  The smooth
increase of the suppression for $\epsilon > \epsilon_D$ in a plasma with radius
$R_{\rm Pb}$ and $p_T^\psi \approx 0$ is shown in the remaining curves.  
All calculations assume $\sigma_{\psi N} = 4.8$ mb and 
$\sigma_{\psi {\rm co}} = 0.67$ mb with eq.~(\ref{survfeed})
modified to include plasma suppression,
\be
S_{AB}(b,s) = 0.58 \, S_\psi^{\rm co} \, S_\psi^{\rm QGP}(\epsilon) + 0.3 \,
S_{\chi_c}^{\rm co} \, S_{\chi_c}^{\rm QGP}(\epsilon) + 0.12 \,
S_{\psi'}^{\rm co} \, S_{\psi'}^{\rm QGP}(\epsilon) \, \, .
\label{feedqgp} \ee

A sudden change of slope in $\psi/$DY appears when the plasma suppression
begins, even without assuming a sharp threshold, not predicted
by hadronic models.
Note that case III overpredicts the suppression and that the thresholds 
introduced in case I are somewhat low relative to
\cite{na50bnl}. Increasing $p_T$ with $R = R_{\rm Pb}$ does not significantly
change the suppression pattern.  If $R = 1$ fm, the plasma contribution is
negligible for $p_T > 4$ GeV in case I and II while case III
is comparable to the dashed and dot-dashed
curves when $p_T \approx 3$ GeV.  Since 
$\langle p_T \rangle \approx 1$ GeV at $\sqrt{s} = 19.4$ GeV \cite{na38pt}, 
the magnitude of the suppression suggests that only the $\chi_c$
and $\psi'$ are suppressed at current energies. Similar results are obtained if
$\sigma_{\psi N} = 7.3$ mb and comovers are not included.  

The $\psi'/\psi$ ratio assuming feeddown is shown in 
Fig.~\ref{figuresepssun}(b).  If a sharp threshold occurs, 
such as in case I,
the $\psi'/\psi$ ratio vanishes above $\epsilon_D^{\psi'}$ and
abruptly increases at $\epsilon_D^{\chi_c}$, an unlikely scenario in a real
collision.  
Case II produces agreement with the data but cannot explain why
the feeddown scenario disagrees with the S+U data.  An enhanced
excitation of $\psi$ into $\psi'$ near the
chiral transition \cite{Shuryak} could also decrease the $\psi'/\psi$
ratio.

\section{Summary} 

We have reviewed $\psi$ suppression by hadronic means and discussed the $E_T$
dependence of a plasma component in Pb+Pb interactions.  We showed that the
naive $A$ dependencies of comover scattering and nuclear absorption 
are identical.  We have extended our comover calculation \cite{GV2}
and found that the S+U and Pb+Pb data are not
simultaneously described in our approach
although agreement in other scenarios with different models of secondary
production cannot be ruled out \cite{Armesto,cassing,capella}.  Thus
$\psi$ suppression by plasma production should also be considered.

The plasma models depend strongly on $\mu(T)$. 
Sharp thresholds do not occur unless introduced, as shown also
in \cite{plasma,kns2}.  In particular,
assuming a central plasma density proportional to $n_{AB}(b,s)$ \cite{plasma},
only an increase in slope occurs at the onset of the 
plasma phase.  The $E_T$ and $\epsilon$ correlation needs
to be clarified, especially since $E_T$ fluctuations cause the $\epsilon$
bins to overlap \cite{na50bnl}.  A positive plasma indication also needs to be
confirmed in other measurements.\\

{\bf Acknowledgments}
I would like to thank B. Chaurand, M. Gonin and L. Kluberg for discussions
about the recent NA50 results and for kind hospitality while visiting the
Ecol\'e Polytechnique.  I would also like to thank S. Gavin for enjoyable
collaboration and J.-P. Blaizot, M. Gyulassy, V. Koch, R. Mattiello,
J.-Y. Ollitrault and H. Satz for discussions.

\end{document}